\documentclass[preprint,showpacs,aps,prb,superscriptaddress]{revtex4-1} 
\usepackage{dcolumn}% Align table columns on decimal point
\usepackage{bm}% bold math
\usepackage[colorlinks,hyperindex, urlcolor=blue, linkcolor=blue,citecolor=black, linkbordercolor={.7 .8 .8}]{hyperref}
\usepackage{graphicx}
\usepackage{tabularx}
\usepackage{amsfonts}
\usepackage{amsmath}
\usepackage{amssymb}
\usepackage{amsbsy}
\usepackage{setspace}
\usepackage{nicefrac}

\newcommand\etal{{\it et~al.}}

\begin{document}

\newcommand{\refs}{/Users/erikku/Documents/Papers2/papers.bib}

\title {Electronic transport and scattering times in tungsten-decorated graphene}

\author{Jamie A. Elias}
\affiliation{Department of Physics, Washington University in St. Louis, 1 Brookings Dr., St. Louis MO 63130, USA}
\author{Erik A. Henriksen}
\affiliation{Department of Physics, Washington University in St. Louis, 1 Brookings Dr., St. Louis MO 63130, USA}
\email{henriksen@wustl.edu}

\begin{abstract}
The electronic transport properties of monolayer graphene have been studied before and after the deposition of a dilute coating of tungsten adatoms on the surface. For coverages up to 2.5\% of a monolayer, we find tungsten adatoms simultaneously donate electrons to graphene and reduce the carrier mobility, impacting the zero- and finite-field transport properties. Two independent transport analyses suggest the adatoms lie nearly 1 nm above the surface. The presence of adatoms is also seen to impact the low field magnetoresistance, altering the signatures of weak localization. 

\end{abstract}

\date{\today}

\maketitle

\section{Introduction}

The two-dimensional electronic system in single layer graphene is inherently unprotected from external influences and thus can be readily altered by proximity to supporting substrates and incidental adsorbates\cite{Schedin:2007vy,Jang:2008eo,Chen:2008tj}. A common and predictable outcome of such interactions is a more disordered electronic system. However there is much interest in the potential to use surface adsorbates to advantageously alter the electronic properties of graphene. A key example is the recent focus on boosting the weak native spin-orbit interaction in graphene in an attempt to engineer topological band structure effects\cite{Weeks:2011cu,Hu266801}. A strong motivation is the desire to realize the Kane-Mele Hamiltonian\cite{Kane:2005hl}, which consists of the relativistic Dirac-like dispersion of graphene plus an intrinsic spin-orbit coupling term that together give rise to a quantum spin Hall insulator. To date, numerous theoretical works address the potential of several different transition metal atoms to play the role of spin-orbit donors leading to graphene-based topologically insulating systems\cite{CastroNeto:2009bg,Ding:2009ky,Qiao:2010cm,Abdelouahed:2010gq,Ding:2011bg,Weeks:2011cu,Hu266801,Jiang:2012io,Zhang:2012kh,Shevtsov:2012is,Ma:2012kl,Chang:2014hg,Hu:2015ha}. Furthermore, there is a wide range of proposals for altering the electronic properties of graphene with adatoms beyond spin-orbit physics, including for the possibility of novel magnetic systems\cite{Sevincli:2008cv,Eelbo:2013ft} and even superconductors\cite{Profeta:2012hg,Ludbrook:2015hs}. 

In this work we explore the effect of tungsten (W) adatoms on the electronic transport of single layer graphene. We find that dilute W coatings cause a significant charge doping along with an increase in scattering and reduced mobility. We experimentally investigate several characteristic scattering times via measurements of the zero-field conductivity, Shubnikov-de Haas oscillations of the magnetoresistance, and low-field signatures of weak localization in magnetoresistance. Our findings are consistent with a picture of isolated W adatoms that become ionized upon donating charge to the graphene, and hence increase the scattering potential experienced by the electrons.

\section{Experiment}

Electronic transport measurements were performed in a cryostat with a 13.5 T superconducting solenoid, using a custom-built sample stage in which graphene samples are mounted facing down toward a small thermal evaporator. Tungsten wires, $20~\mu$m diameter and of 99.95\% purity, are located approximately 8 cm below the sample for use as evaporation sources. The evaporation rate of W atoms is controlled by passing a current through the wire while simultaneously monitoring any changes that occur in the electronic transport of the sample. During evaporation, the sample temperature rises to approximately 40 K while the rest of the cryostat remains close to 4 K. The density of deposited atoms may be estimated from changes to the graphene transport caused by charge doping from the adatoms as discussed below, and independently by measuring the change in diameter of the tungsten wire sources. Graphene samples are produced starting with mechanical exfoliation of Kish graphite onto Si wafers having a 300 nm thermal oxide, followed by fabrication of electrical contacts by electron beam lithography and thin film Cr/Au deposition. Transport data is acquired before and after evaporation using standard low-frequency AC lock-in techniques. Applying a gate voltage, $V_g$, to the degenerately-doped Si substrate allows control of the free carrier density in graphene, $n=\alpha (V_g - V_{g0})$, where the coefficient $\alpha = 7.0\times10^{10}$ cm$^{-2}$V$^{-1}$ is determined from oscillations in the magnetoresistance at high fields. Here we present results from a single layer graphene sample etched by an O$_2$ plasma into a 2-micron-wide Hall bar (see inset to Fig.\ref{cond}(a)); similar behavior has been observed in a second sample.

The deposition of tungsten atoms on to the surface of graphene impacts the electronic transport in several ways as will be presented below. An independent measurement of the deposited adatom density is desirable. Here we estimate the adatom density by measuring the diameter of the tungsten wire source in a scanning electron microscope both before and after the experiment. Geometry then enables an estimate of the adatom density. This method gives the \emph{final total} density deposited by all three evaporations performed. This density is found to be $5\times10^{13}$ cm$^{-2}$, covering 2.5\% of the unit cells in graphene (or just over 1 W atom per 100 C atoms). We estimate the uncertainty in this value to be $\pm20$\%.

\section{Results}
\subsection{Transport at zero magnetic field}

\begin{figure}[t]
\includegraphics[width=0.5\textwidth]{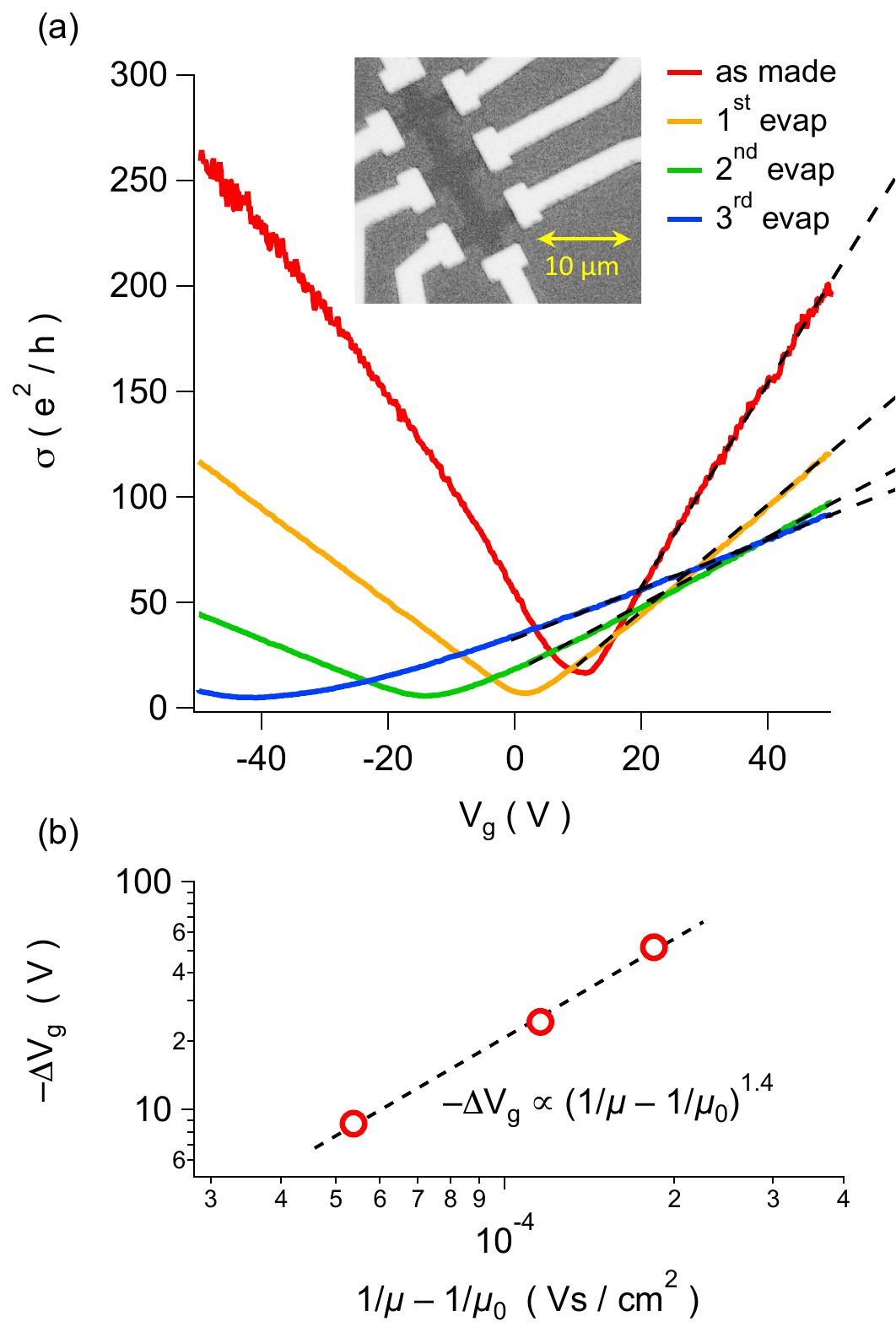}
\caption{(Color online) (a) Conductivity, $\sigma$, vs.~gate voltage, $V_g$, for the monolayer graphene Hall bar sample shown inset to the figure. The as-made sample has a conductivity minimum at $V_g = 10$ V, which shifts to the left indicating electron doping after each of three subsequent tungsten evaporations. Black dashed lines show linear fits used to extract the field-effect carrier mobility. (b) The gate voltage shift, $\Delta V_g = V_{g,min} - V_{g,0}$, of the minimum conductivity in part (a) for each evaporation, plotted against the change in the inverse mobility, where $\mu_0$ is the mobility of the as-made trace (red curve in part (a) ). The shift has a power law dependence on the inverse mobility with slope of 1.4, consistent with point-like scattering\cite{McCreary:2010hr,Chen:2008tj,Adam:2007wa}. \label{cond}} 
\end{figure}

Figure \ref{cond}(a) shows the measured conductivity as a function of gate voltage for the sample used in this study, starting with data obtained from the as-made device (red trace) and continuing with traces recorded after three successive depositions of tungsten atoms, shown as the orange, green, and blue curves. With $n \propto V_g$, each trace exhibits a linear dependence on carrier density away from the conductivity minimum. This implies a constant carrier mobility, $\mu = (1/e) d\sigma / dn$, that is observed to decrease after each evaporation. Moreover the charge neutrality point at the conductivity minimum of each curve is seen to shift to the left indicating electron doping of the graphene; and the curvature of the conductivity minimum broadens, suggesting an increase in carrier density inhomogeneity\cite{Chandni:2015kv}. Broadly, these observations are consistent with prior works on the impact of potassium adatoms on graphene\cite{Chen:2008tj,Yan:2011bg}; Ti, Fe, and Pt adatoms\cite{Pi:2009co}; Au adatoms\cite{McCreary:2010hr}; indium\cite{Chandni:2015kv,Jia:2015dc}; and iridium as well\cite{Wang:2015cm}; and altogether strongly suggest that W adatoms donate electrons to graphene, becoming ionized impurities that enhance the scattering and reduce the mobility. 

Away from the minimum in conductivity at charge neutrality, the conductivity of graphene can be written as\cite{Nomura:2006bi,2006JPSJ...75g4716A,Adam:2007wa} 
\begin{equation}
\sigma^{-1}(n)~=~\sigma^{-1}_{ci}(n)~+~\sigma^{-1}_{sr}~,\nonumber
\end{equation}
reflecting two sources of scattering: screened charged impurities and short-ranged scattering as might arise at edges or vacancies. In the former case the conductivity is linear in the carrier density, $\sigma_{ci}(n) = C |n|/n_{imp}$ where $n_{imp}$ is the impurity density and $C$ has been theoretically calculated\cite{Adam:2007wa} to be $C\approx20~ e^2 / h$ in the limit that the charged impurities lie in the graphene plane. Short-ranged impurities, on the other hand, lead to a conductivity that is independent of density. In many graphene-on-SiO$_2$ devices including those used in this work, $\sigma_{sr}>>\sigma_{ci}$ and thus the conductivity is simply observed to be linear in density.

The conductivity data allow us to extract the added charge density $\Delta n = \alpha \Delta V_g$ donated by the adatoms by measuring the shift of the conductivity minimum along the gate voltage axis, and also the change in the impurity density $\Delta n_{imp} = (C/e)(1/\mu - 1/\mu_0)$, where we subtract off the contribution of the initial impurity distribution in the as-made sample. Naively, one might expect each adatom to donate one electron such that $\Delta n = \Delta n_{imp}$. However, we instead find the induced charge density to be $2-3$ times larger than the $n_{imp}$ values extracted from the conductivity data of Fig.~\ref{cond}(a). We note that to determine $n_{imp}$ we use the aforementioned value $C=20~e^2/h$ which is strictly true only for $z=0$, where $z$ is the effective height of the impurities above the graphene plane. In fact the self-consistent theory of Adam \etal~predicts\cite{Adam:2007wa} that $C$ has a super-linear density dependence for $z>0$, growing increasingly with $z$. Without \emph{a priori} knowing the height of the adatoms above the graphene, we can make an estimate by first \emph{assuming} that the increase in $n_{imp}$ is in fact \emph{equal} to $\Delta n$, and then applying this theory to calculate the value of $z$. The fact that $C$ increases with $z$ explains why $n_{imp}$ is initially underestimated. The results of our analysis are given in Table I where we list the observed change in density, $\Delta n$, the $n_{imp}$ values calculated from the slope $d\sigma/dn$ using $C=20~e^2/h$, and the calculated heights, $z$. With increasing W coverage, $z$ becomes close to 1 nm. Although this is greater than the $z = 0.16-0.17$ nm separation predicted for W atoms above the center of a graphene honeycomb by density functional theory (DFT) calculations\cite{Nakada:2011fr,Zhang:2012kh}, it is not altogether unreasonable. We further note that predictions for the charge transfer from W to graphene is range between 0.56 and 0.93 electrons/atom\cite{Nakada:2011vs,Manade:2015em}; the average of these would increase the calculated height by roughly 30\%.

In Figure \ref{cond}(b) we plot the shift in gate voltage, $-\Delta V_g$, vs.~the change in the inverse mobility, $1/\mu - 1/\mu_0$. This quantity is both predicted and experimentally found to obey a power law, $-\Delta V_g \propto (1/\mu - 1/\mu_0)^b$, where $b$ is typically $1.2-1.3$ for point-like charged impurities (as opposed to clusters, for which $b<1$)\cite{McCreary:2010hr,Chen:2008tj,Adam:2007wa}. We find $b=1.4$. Thus altogether the zero-field conductivity implies that W adatoms are isolated, charged impurities lying approximately 1 nm above the surface.

\begin{figure*}[t!]
\includegraphics[width=\textwidth]{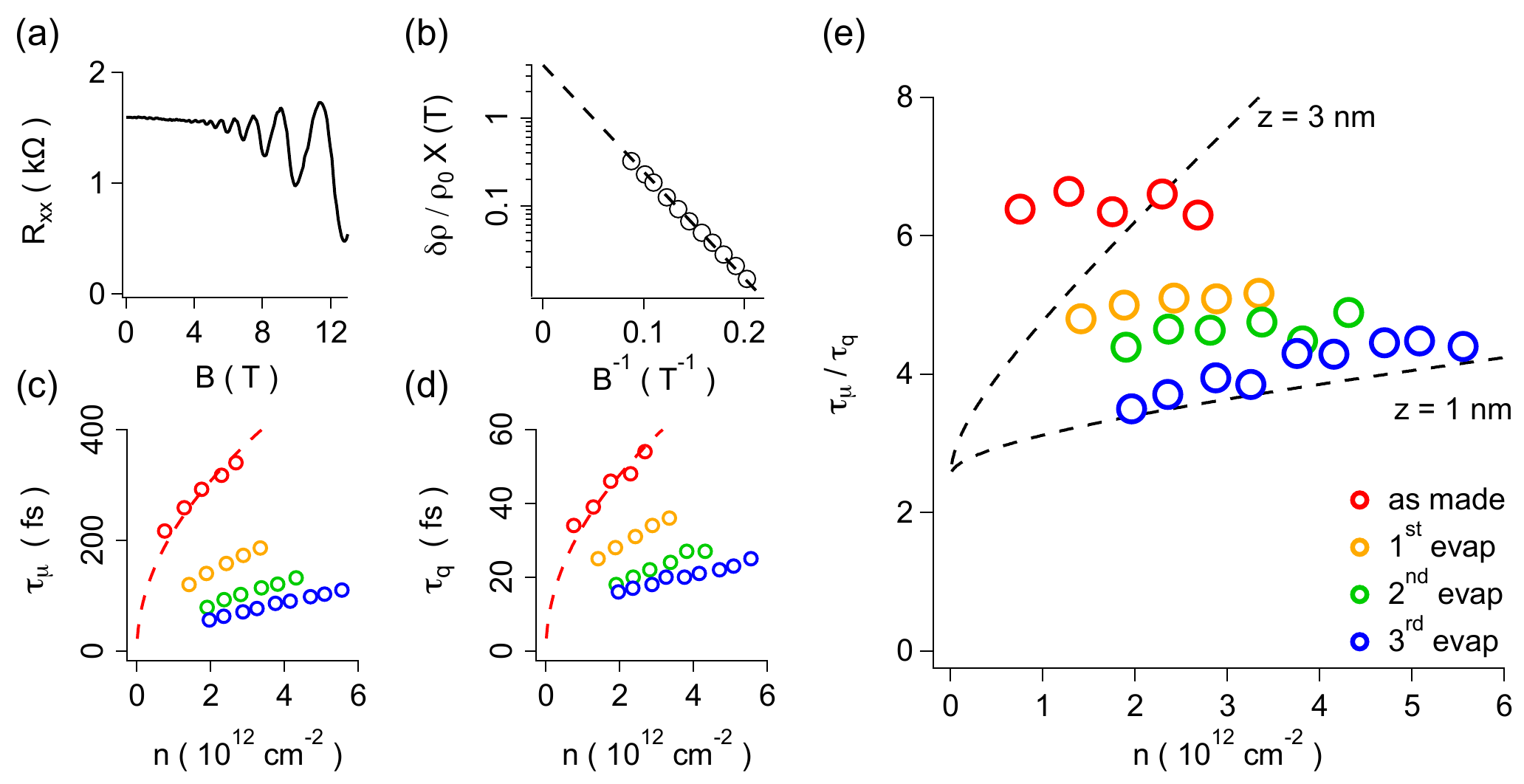}
\caption{(Color online) (a) representative Shubnikov-de Haas trace after the second W evaporation, at a density $n = 4.3\times 10^{12}$ cm$^{-2}$. (b) The corresponding analysis of the temperature-corrected oscillation amplitudes vs.~$B^{-1}$ after Coleridge\cite{Coleridge:1991ui} and Hong\cite{Hong:2009cj}; the linear fit is constrained to a $y$-intercept of 4, and the slope is proportional to the quantum scattering rate $\tau_q^{-1}$. (c),(d) Transport and quantum scattering times extracted from conductivity data of Fig.~\ref{cond} and analysis of SdH traces, respectively. Red dashed lines are fits following $\sqrt{n}$. (e) Ratio of transport to quantum scattering time, $\tau_{\mu} / \tau_q$, vs.~carrier density, $n$, for the as made sample and after each evaporation. Uncertainty in the data is given by the symbol size. The dashed black lines are calculated using the theory of Hwang \& das Sarma\cite{Hwang:2008gg}.\label{ratio}} 
\end{figure*}

\subsection{Comparison of transport and quantum scattering times}

We now investigate the behavior of the transport and quantum scattering times, $\tau_{\mu}$ and $\tau_q$, as W atoms are deposited on graphene. Both parameters are a measure of electron scattering, but where the single particle relaxation (or ``quantum'') time $\tau_q$ is sensitive to \emph{all} scattering events, the transport time $\tau_{\mu}$ only measures those that contribute to the resistance of the material, e.g. forward scattering processes are ignored. In standard 2D systems, backscattering (where $|{\bf k}_{final}-{\bf k}_{initial}| = 2k_F$) is the most efficient at limiting $\tau_{\mu}$, but these events are suppressed in single layer graphene\cite{Ando:1998wr} leaving ``right-angle'' scattering to have the strongest impact on the transport time. The scattering rates are found by integrating over the total angular scattering potential $Q(\theta)$ as
\begin{gather}
\frac{1}{\tau_q} = \int_0^{\pi} ~Q(\theta)~(1+{\rm cos} ~\theta)~d\theta, \nonumber\\
\frac{1}{\tau_{\mu}} = \int_0^{\pi} ~Q(\theta)~(1-{\rm cos}^2 ~\theta)~d\theta,\nonumber
\end{gather}
\noindent where factors of $1+{\rm cos} ~\theta$ in each formula account for the suppression of $2k_F$ scattering, and the additional factor of $1-{\rm cos} ~\theta$ in the transport scattering rate limits the effect of forward scattering.

The ratio $\tau_{\mu} / \tau_q$ can be used to discriminate between the type and location of scattering potentials, for instance short-range ($\delta$-function) impurities scatter equally into all angles and thus $\tau_{\mu} = 2 \tau_q$ where the factor of 2 is linked to the absence of backscattering, while Coulomb scattering leads to an increase in forward scattering events due to its long-ranged nature\cite{DasSarma:1985uw,Hwang:2008gg}, so that $\tau_{\mu} / \tau_q > 2$. Indeed in high mobility GaAs 2D systems $\tau_{\mu} / \tau_q$ can exceed 100 due to the exceptional purity of the host crystal and the fact that ionized impurities are removed many 10s of nm from the 2D layer.

In graphene-on-SiO$_2$ this ratio is expected to be small due to strong scattering caused by close coupling of the graphene sheet to the substrate, as indeed was observed by Hong \etal\cite{Hong:2009cj}. In particular, theoretical work predicts that $\tau_{\mu} / \tau_q <2$ for short-ranged scattering, $\tau_{\mu} / \tau_q >2$ for (screened) Coulomb scattering when the impurities lie in the plane, and becomes increasingly larger for charged impurities that are set back a distance $z$ above the plane\cite{Hwang:2008gg}. The reason is that the more distant a Coulomb scatterer is, the smaller the scattering angle will be which preferentially limits the quantum scattering time. Moreover, the ratio depends on whether impurities are either isolated or clumped together in clusters, in which case the charge doping efficiency and hence the number of ionized scatterers is reduced\cite{McCreary:2010hr}. Indeed, for clusters the ratio $\tau_{\mu} / \tau_q$ is predicted to increase by roughly the number of impurities per cluster as the total scattering cross-section outstrips the rate of backscattering\cite{Katsnelson:2009ja}. To learn more about the impact of W adatoms, we have studied the $\tau_{\mu} / \tau_q$ ratio as it is impacted by tungsten adatoms. We extract the transport scattering time from the conductivity data of Fig.~\ref{cond}, and the $\tau_q$ values from an analysis of Shubnikov-de Haas (SdH) oscillations of the magnetoresistance at high magnetic fields and over a range of carrier densities for electron-doped graphene, in the as-made sample and following each evaporation. The amplitude of Shubnikov-de Haas oscillations is generally well-described by the first term of the Lifshitz-Kosevich equation\cite{Gusynin:2005ht},
\begin{equation}
\frac{\delta \rho_{xx}}{\rho_0} = 4 ~X_{th} ~{\rm exp}\left(-\frac{\pi}{\omega_c \tau_q} \right); ~~X_{th} = \frac {2 \pi^2 k_B T / \hbar \omega_c}{{\rm sinh}(2 \pi^2 k_B T / \hbar \omega_c)} \nonumber
\end{equation}
\noindent where $\rho_0 = \rho(B=0)$ and $\omega_c = e B / m^*$ is the cyclotron frequency with the effective mass $m^* = \hbar \sqrt{\pi n} / v_F$.

Figure \ref{ratio}(a) shows a representative SdH trace at a density $n = 4.3 \times 10^{12}$ cm$^{-2}$ after the second tungsten evaporation. The logarithm of the amplitude of the oscillations, divided by $\rho_0$ and the thermal damping factor $X_{th}$, yields a straight line when plotted vs.~$1/B$ as shown in Fig.~\ref{ratio}(b), with a slope that is inversely proportional to the quantum scattering time\cite{Coleridge:1991ui,Hong:2009cj}. The transport and quantum scattering times we find are plotted in Fig.~\ref{ratio}(c) and (d), respectively. Both follow a roughly $\sqrt{n}$ dependence, shown by the dashed red curves.

\begin{table}[b]
\caption{Adatom-induced electron density $\Delta n$ determined from shift in minimum conductivity; the impurity density $n_{imp} = (C/e)(1/\mu - 1/\mu_0)$; and the height of impurities above the plane, $z$, in nm, calculated using the theory of Adam \etal~as discussed in the text\cite{Adam:2007wa}.}
\begin{tabular}{|c|c|c|c|}
\hline
State of sample & $\Delta n$ (cm$^{-2}$) & $n_{imp}$ (cm$^{-2}$) & $z$ (nm) \\
\hline
as made & 0 & $2.9\times10^{11}$ & 0.08 \\
1$^{st}$ evap & $6.1\times10^{11}$ & $5.5\times10^{11}$ & 0.6 \\
2$^{nd}$ evap & $1.7\times10^{12}$ & $8.5\times10^{11}$ & 0.92 \\
3$^{rd}$ evap & $3.6\times10^{12}$ & $1.2\times10^{12}$ & 1.1 \\
\hline
\end{tabular}
\end{table}

In Fig.~\ref{ratio}(e) we plot the ratio of these scattering times as a function of carrier density, for the as-made sample and following each evaporation. A clear downward trend is visible with the ratio dropping from $6-7$ down to $3-4$ over the course of the depositions, and while the ratio is more or less constant in the as-made sample, with each evaporation a slight but clear increase in the slope emerges. The density range explored is limited by a combination of the shift in the minimum conductivity due to electron doping, and the associated decrease in mobility which smears out SdH oscillations at lower densities. We compare these data to predictions from the theory of Hwang \& das Sarma\cite{Hwang:2008gg} for the variation of $\tau_{\mu} / \tau_q$ with the effective height $z$ of charged impurities above the graphene sheet, for $z = 1$ and $3$ nm. Assuming that charged impurity scattering in the as-made sample arises from oxide charges and dangling bonds in the underlying SiO$_2$ surface, 3 nm is a reasonable value given the few-nm RMS surface roughness of SiO$_2$, and is close to the 2 nm found previously for graphene-on-SiO$_2$ devices\cite{Hong:2009cj,Romero:2008dra}. The decrease to a 1 nm separation after the evaporations implies that ionized W adatoms lie closer to the surface, and agrees with the value found above from consideration of the zero-field conductivity. As previously noted, however, the separation distance for tungsten adatoms predicted by DFT calculations is rather smaller, $z = 0.16-0.17$ nm\cite{Nakada:2011fr,Zhang:2012kh}. A similar discrepancy in separation distances was found for indium adatoms\cite{Chandni:2015kv}.

Finally we note that although the ending impurity-graphene separation of 1 nm found from the scattering time ratio agrees with the value found from the zero-field conductivity analysis above, the \emph{initial} separations prior to any W deposition are in sharp disagreement. This may speak to our ignorance of the impurity distribution in the as-made sample, in which scattering sources other than charged impurities will play a larger role. For instance, we have not isolated the contribution of long- and short-ranged scattering in our analysis of the zero-field conductivity, as in all cases the charged impurity linear-in-$n$ scattering dominates.

\begin{figure}
\includegraphics[width=0.5\textwidth]{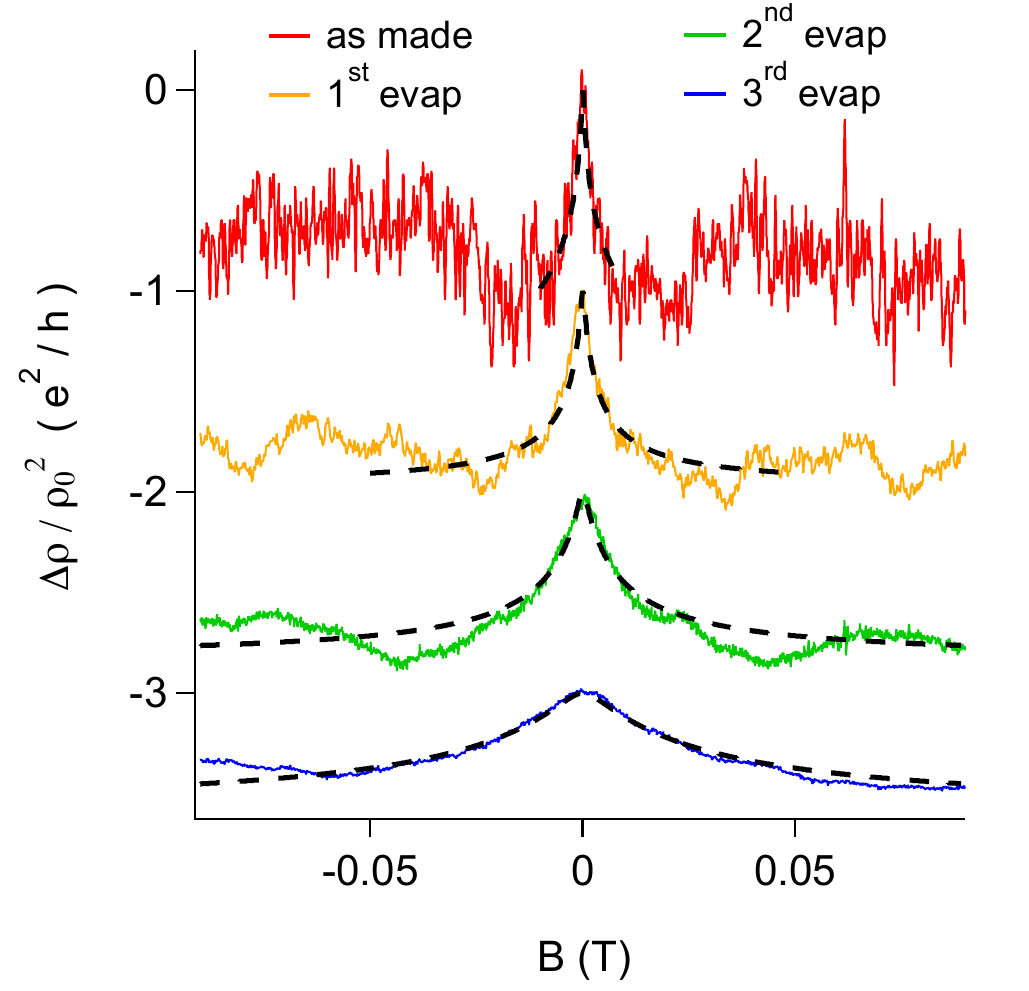}
\caption{(Color online) Low-field magnetoresistance for a carrier density $n=1.4\times 10^{12}$ cm$^{-2}$, showing characteristic weak localization peaks about $B=0$. Traces are vertically offset for clarity. The dashed black lines are fits using Eq.~\ref{wleq}, with the field range restricted to the diffusive regime $(l_{\mu} / l_B)^2<<1$; the fitting parameters are plotted in Fig.~\ref{taus}. \label{fits}} 
\end{figure}

\subsection{Weak localization}

At magnetic fields below 50 mT, the sample shows clear signs of weak localization in the magnetoresistance. Figure \ref{fits} shows four traces, plotted as $\delta \rho / \rho_0^2$ where $\delta \rho = \rho(B) - \rho_0$, for the as-made sample and following each evaporation at a carrier density of $n=1.4\times 10^{12}$ cm$^{-2}$. All four traces show a narrow peak that is roughly $e^2/h$ in magnitude, along with universal conductance fluctuations that are symmetric in the field. With each evaporation both the localization peak and the conductance fluctuation features are seen to broaden and become reduced in amplitude.

\begin{figure}
\includegraphics[width=0.5\textwidth]{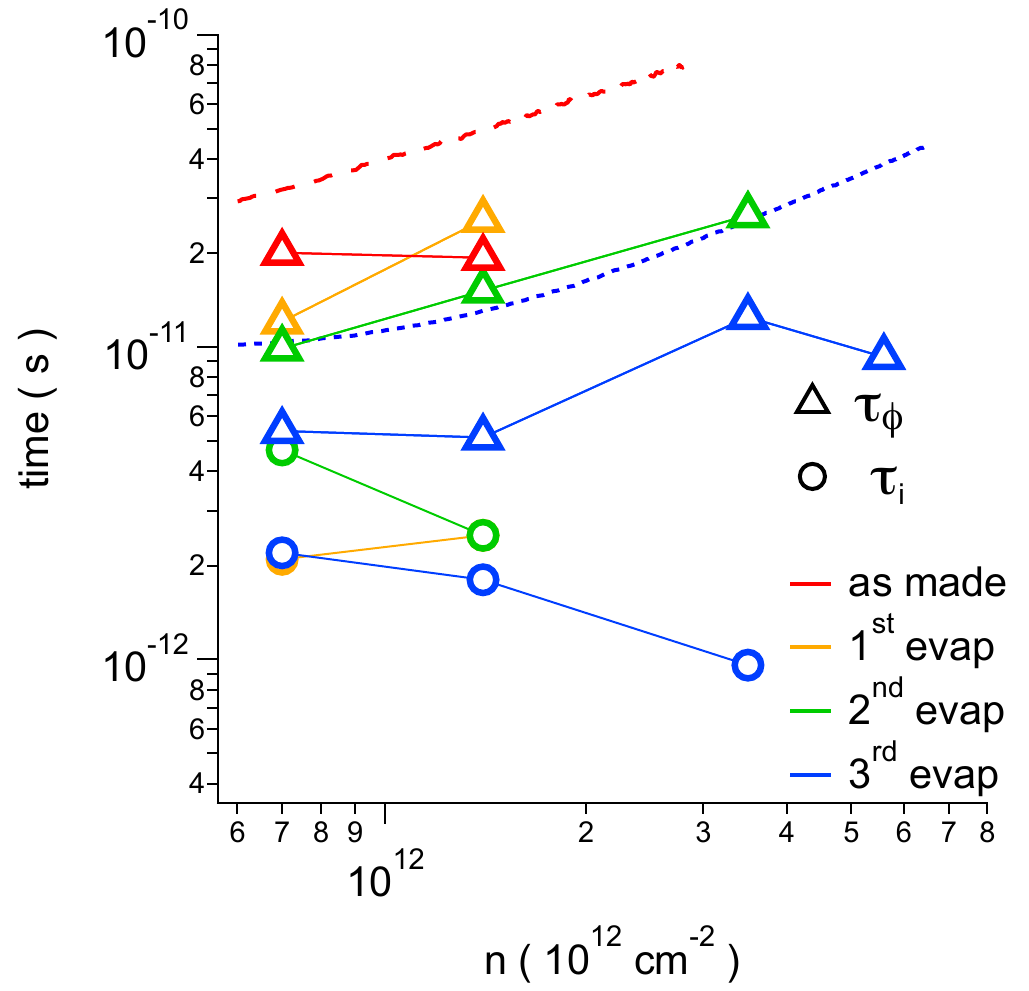}
\caption{(Color online) Phase-breaking time, $\tau_{\phi}$, and intervalley scattering time, $\tau_i$, extracted from curve fits to the low-field magnetoresistance using Eq.~\ref{wleq}, as illustrated in Fig.~\ref{fits}. The theoretical phase breaking times calculated with Eq.~\ref{tauphi} are shown as lines in the upper portion of the figure, using data from the as-made sample (dashed red) and after the 3$^{rd}$ evaporation (dotted blue). Trends in the data reflect the increase (decrease) in conductivity with carrier density (successive evaporations).\label{taus}} 
\end{figure}

Analysis of the localization correction to the conductivity can yield useful information on characteristic scattering times including the phase coherence time $\tau_{\phi}$ and various other scattering mechanisms\cite{Bergmann:1984ty}. In graphene these may include inter-valley scattering rates, intra-valley scattering processes including sublattice symmetry effects and trigonal warping of the Dirac cones, and spin-orbit effects\cite{Mccann:2006tz,Mccann:2012hd}. We perform fits to our data using a simplified version of the theory developed by McCann \& Fal'ko \etal~which ignores small corrections to the weak localization due to intra-valley scattering processes\cite{Falko:2007du}:
\begin{gather}
\frac{\Delta \rho}{\rho_0^2} = - \frac{e^2}{\pi h} \left[ F \left(\frac{B}{B_{\phi}}\right) - F \left(\frac{B}{B_{\phi}+2B_i} \right) \right], \label{wleq}\\
F(z)={\rm ln}(z)+\psi \left(\frac12+\frac1z \right), ~~B_{\phi,i}=\frac{\hbar}{4De}\tau^{-1}_{\phi,i}. \nonumber
\end{gather}
\noindent Here $D=v_F^2\tau_{\mu}/2$ is the diffusivity and $\psi$ is the digamma function. Use of this simplified theory is justified by the fact that graphene-on-SiO$_2$ samples have a high intra-valley scattering rate that results in a negligible contribution to localization effects\cite{Tikhonenko:2008ec,Couto:2014ip,Falko:2007du}. Indeed in comparing fits to our data made using either Eq.~\ref{wleq} or the full WL expression for graphene\cite{Mccann:2006tz} we find virtually no difference in the fitting curves. However, the inclusion of an additional fitting parameter for intra-valley scattering in the full theory leads to $\chi^2$ values that are poorly constrained and large uncertainties in the scattering times. Thus we obtain our numerical estimates of $\tau_{\phi}$ and $\tau_i$ from fits using Eq.~\ref{wleq}, applied over a magnetic field range such that the elastic mean free path is much less than the magnetic length\cite{Lee:1985tl}, $l_{el}^2 << l_B^2$, where $l_B = \sqrt{\hbar / e B}$.

Figure \ref{taus} shows the results of our fitting procedure for values of the dephasing time $\tau_{\phi}$ and the inter-valley scattering time $\tau_i$. The differing density ranges used for the as-made sample and each separate evaporation are a consequence of the electron doping which, for the fixed gate voltage range employed, accesses an enlarged span of electron densities with each successive deposition. Not shown in Fig.~\ref{taus} is the distribution of transport scattering times, but even the largest $\tau_{\mu}$ measured-- in the as-made sample at the highest density explored-- is only 0.3 ps so that in all cases $\tau_i$ exceeds $\tau_{\mu}$ by at least one or two orders of magnitude. Roughly speaking, we find the dephasing times $\tau_{\phi}$ show a modest increase with carrier density and a decrease with each tungsten deposition. Such behavior is in accordance with the predictions of Altshuler, Aronov and Khmel'nitski\cite{Altshuler:1982ue}:
\begin{equation}
\tau_{\phi}^{-1}=\frac{k_BT}{2 \hbar}\frac{{\rm ln}(\pi\hbar \nu D)}{\pi\hbar \nu D} = \frac{k_B T}{2 \hbar}\frac{{\rm ln}(k_F l)}{k_F l}
\label{tauphi}
\end{equation}
with $D$ again the diffusivity and $\nu$ the density of states at the Fermi level. The expression on the right includes the single layer graphene density of states $\nu = 2 E_F / (\pi \hbar^2 v_F^2)$. This formula is plotted in Fig.~\ref{taus} using the transport parameters for the as-made sample (red dashed line) and after the final tungsten evaporation (blue dotted line). The data clearly follow the general trend illustrated by these curves. Meanwhile the inter-valley scattering times, although suffering from a fair degree of scatter, do tend to show a decrease with each evaporation. This is surprising: if tungsten adatoms act as charged impurities only intra-valley scattering should increase, and an increase in $\tau_i$ would be expected due to the accompanying reduction in diffusivity. Moreover the inter-valley times $\tau_i \approx 1-5$ ps correspond to a scattering lengths $l_i = \sqrt{D \tau_i}~=~200-300$ nm, rather smaller than the $2~\mu$m width of the device, although device edges are where the strongest inter-valley scattering is expected. Tungsten atoms have a strong inherent spin-orbit coupling which is predicted\cite{Zhang:2012kh} to be inherited by graphene. Thus arguably we should study the low-field magnetoresistance with curve fits to the theory that incorporates the physics of spin-orbit scattering\cite{Mccann:2012hd}. We have attempted this and find the fits to be generally inferior to those found using Eq.~\ref{wleq}. Additionally, it is clear that no obvious signatures of spin-orbit coupling, e.g.~weak anti-localization\cite{1982PhRvL..48.1046B,Mccann:2012hd}, are observed. Altogether these findings suggest that any scattering due to an inherited spin-orbit interaction, if present at all, is weaker than inter-valley scattering.

\section{Discussion}

In this work we present the results of electronic transport measurements on a sample of monolayer graphene with an increasing, but always small, density of tungsten adatoms. Measurements of the zero-field conductivity, the ratio of transport and quantum scattering times, and weak localization in the magnetoresistance point toward a now-familiar picture\cite{Chen:2008tj,Yan:2011bg,Pi:2009co,McCreary:2010hr,Chandni:2015kv,Jia:2015dc,Wang:2015cm} for metal adatoms wherein isolated impurities donate charge to the graphene substrate and become ionized impurities with a concomitant increase in scattering and reduction of the mobility. However there is one standout feature in our experiment, namely as mentioned above we estimate that over 10 times as many W atoms were deposited as can be accounted for by magnitude of electron doping measured by the shift in the minimum conductivity, under the assumption that each atom donates one electron. Relaxing this assumption to within the range of theoretically-predicted values of $0.56-0.93$ electrons/atom\cite{Nakada:2011vs,Manade:2015em} does little to improve the match.

Regarding the W deposition, two scenarios immediately suggest themselves: first, the adatoms may form clusters, or second, fabrication residues may prevent many atoms from reaching the surface. In the former case, we note the migration energy for W adatoms (the difference between binding energies at three high-symmetric sites: atop a C atom, astride a C-C bond, or in the middle of a hexagon) is calculated\cite{Nakada:2011vs} by DFT to be $E_m=1.2$ eV. The hopping rate for migrating adatoms is given by\cite{Zangwill:1988ue} $\nu=(k_B T / h)~{\rm exp}(-e E_m / k_B T)$, which for all practical purposes vanishes below 150 K for W atoms. For completeness we note this strongly disagrees with a second DFT study\cite{Manade:2015em} that finds the diffusion energy to be a mere 20 meV with the stable site to be the C-C bond; clearly this would change our expectations. In any event these considerations apply to atoms already on the surface, however adatoms with a high kinetic energy freshly evaporated from the $\approx 2800$ K hot wire may diffuse on the graphene surface before losing their energy via thermal radiation or phonon emission, and thus have the opportunity to form clusters. Clusters of metal atoms on graphene are known to be far less efficient at charge doping\cite{McCreary:2010hr,Wang:2015cm} and consequently have a limited impact on the mobility. However, the total scattering cross-section remains large for clusters and indeed a key prediction\cite{Katsnelson:2009ja} of cluster-dominated transport is a ratio $\tau_{\mu}/\tau_q >> 1$, precisely the opposite of what we observe.

Thus we consider a second possibility, that fabrication residues prevent most of the evaporated atoms from reaching the surface. The surface of graphene after standard fabrication procedures employing PMMA as a resist for electron-beam lithography has been directly imaged by transmission electron microscopy\cite{Lin:2011fn,Lin:2012bf}, and a thin ($\sim$ nm) coating of PMMA molecules is found to remain even after aggressive thermal annealing procedures (which were not performed on our sample). The images reveal that a significant portion of the surface may be covered by this remnant PMMA, enough to perhaps account for the discrepancy between the expected density of W atoms based on the change in the source wire diameter, and the maximum possible density assuming each atom donates one electron. This PMMA layer also provides a natural means to prevent clustering: if any hot tungsten atom does migrate upon landing, it will shortly encounter this impurity layer and come to a halt.

Finally, we note an induced spin-orbit coupling in graphene has been found by measuring so-called ``non-local'' voltages, by driving a current through one region of a modified (e.g. hydrogenated) graphene sample and finding a voltage drop far away where no appreciable current flow is expected in an ohmic material\cite{Balakrishnan:2014jg}. We have checked for such non-local voltages and find results consistent with zero.

\section{Conclusion}

In conclusion, we have performed a transport study of graphene with a dilute coating of W adatoms. The adatoms induce an electron doping of the graphene and a reduction of the mobility, enforcing a linear dependence of the conductivity on density consistent with charged impurity scattering. Analysis of the changes in the conductivity suggest the W atoms reside approximately 1 nm above the surface. Similar to the case of indium adatoms\cite{Chandni:2015kv}, this distance is unexpectedly large given that the atoms clearly are close enough to donate charge, and it also disagrees with the results of DFT calculations. One possibility is the height discussed in the self-consistent theory should be considered an ``effective'' distance that is correlated with, if not identical to, the actual physical separation that DFT attempts to calculate. This clearly requires further experimental investigation, preferably by a scanned probe method that is sensitive to the adatom height.

We have also performed a study of the ratio of the transport to quantum scattering time, $\tau_{\mu} / \tau_q$, finding the ratio to decrease from $6-7$ down to $3-4$ as the density of W adatoms increases. The adatom height inferred by comparison to theoretical calculations is approximately 1 nm, in agreement with the estimate found from the zero-field conductivity. 

Clear signatures of weak localization are seen at low magnetic fields. The dephasing times extracted from fits are in agreement with values expected from theory. The intervalley scattering times are shorter than expected, but the precise scattering potential of the as-made device is not known and there may be defects or surface impurities that cause scattering with large momentum transfers. All transport data point to a picture wherein W deposition at low coverages leads to isolated charged impurities. No evidence of spin-orbit coupling transferred from the W adatoms to the graphene is found.

\begin{acknowledgements}
We thank S.~Adam, J.~Alicea, U Chandni, S.~Das Sarma, M.~Franz, M.~Fuhrer, R.~Mong, J.~Pollanen, and M.-F.~Tu for helpful correspondence and discussions.
\end{acknowledgements}

\bibliography{tungsten}

\end{document}